\documentclass[10pt, twocolumn, pre, aps, superscriptaddress, showpacs]{revtex4-1}
\usepackage{amsmath, graphicx, subfigure, tikz}
\begin{document}

\title{Nematic Ordering in the Heisenberg Spin-Glass System in d=3 Dimensions}

\author{Egemen Tunca}
    \affiliation{TEBIP High Performers Program, Board of Higher Education of Turkey, Istanbul University, Fatih, Istanbul 34452, Turkey}
\author{A. Nihat Berker}
    \affiliation{Faculty of Engineering and Natural Sciences, Kadir Has University, Cibali, Istanbul 34083, Turkey}
    \affiliation{T\"UBITAK Research Institute for Fundamental Sciences, Gebze, Kocaeli 41470, Turkey}
    \affiliation{Department of Physics, Massachusetts Institute of Technology, Cambridge, Massachusetts 02139, USA}

\begin{abstract}
Nematic ordering, where the spins globally align along a spontaneously chosen axis irrespective of direction, occurs in spin-glass systems of classical Heisenberg spins in $d=3$. In this system where the nearest-neighbor interactions are quenched randomly ferromagnetic or antiferromagnetic, instead of the locally randomly ordered spin-glass phase, the system orders globally as a nematic phase.  The system is solved exactly on a hierarchical lattice and, equivalently, Migdal-Kadanoff approximately on a cubic lattice.  The global phase diagram is calculated, exhibiting this nematic phase, and ferromagnetic, antiferromagnetic, disordered phases.  The nematic phase of the classical Heisenberg spin-glass system is also found in other dimensions $d>2$: We calculate nematic transition temperatures in 24 dimensions in $2<d\leq4.$
\end{abstract}
\maketitle

\section{Introduction: Nematic Order Due to Quenched Randomness}

Spin glasses, broadly defined as systems with frozen (quenched) disorder that have locally annulling interactions (frustration), present complex systems with a plethora of distinctive characteristics.  These distinctions include the spin-glass phase and its signature: chaos under repeated scale changes \cite{McKayChaos,McKayChaos2,BerkerMcKay}.  The fractal spectrum of spin-glass chaos has recently been shown to be used as a classification and clustering tool for the broadest of complex data, including multigeographic multicultural music and brain signals.\cite{classif}  The ordering of the spin-glass phase has local fixation within spatial non-uniformity, the direction and magnitude of the local magnetization varying between neighboring points of a lattice, but the direction of local magnetization being firmly fixed relative to the local magnetizations of the neighbors.

The above discussion has been in terms of Ising spins, namely one-component spins, on which the preponderance of spin-glass research has been done.  We find here that for three-component Heisenberg spins, the new ordering evades the directional fixation: the spins globally align along a spontaneously chosen axis irrespective of direction, thus creating a nematic spin phase.  Thus, symmetry is globally broken by the spontaneous choice of a spin axis, but all local magnetizations are zero.

\section{The Model and the General Method}

\begin{figure}[ht!]
\centering
\includegraphics[scale=1]{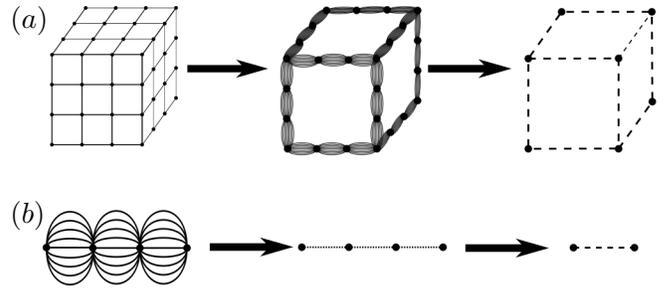}
\caption{(a) Migdal-Kadanoff approximate renormalization-group transformation for the $d=3$ cubic lattice with the length-rescaling factor of $b=3$. In this intuitive approximation, bond moving is followed by decimation. (b) Exact renormalization-group transformation of the $d=3, b=3$ hierarchical lattice for which the Migdal-Kadanoff renormalization-group recursion relations are exact. The construction of a hierarchical lattice proceeds in the opposite direction of its renormalization-group solution. From \cite{BerkerOstlund,Caglar2}.}
\end{figure}

The classical Heisenberg spin-glass system is defined by the Hamiltonian
\begin{equation}
- \beta {\cal H} = \sum_{\left<ij\right>} \, J_{ij} \, \vec s_i\cdot \vec s_j,
\end{equation}
where $\beta=1/k_{B}T$, $J_{ij}=+|J|$ or $-|J|$ (ferromagnetic or antiferromagnetic) with probability $p$ and $1-p$ respectively, the classical spin $\vec s_i$ is the unit spherical vector at lattice site $i$, and the sum is over all nearest-neighbor pairs of sites.

We solve the classical Heisenberg spin glass by a renormalization-group transformation that is exact on the $d=3$ hierarchical lattice and, equivalently, Migdal-Kadanoff approximate \cite{Migdal,Kadanoff} on the $d=3$ cubic lattice (Fig. 1).  The latter much-used approximation is physically intuitive: In a hypercubic lattice where an exact renormalization-group transformation cannot be applied, as an approximation some of the bonds are removed, which weakens the connectivity of the system and, to compensate, for every bond removed, a bond is added to the remaining bonds.  This step is the bond-moving step and constitutes the approximate step of the renormalization-group transformation.  At this point, the intermediate sites can be eliminated by an exact integration over their spin values in the partition function, which yields the renormalized interaction between the remaining sites.  This is called the (exact) decimation step and completes the renormalization-group transformation. As shown in Fig. 1, the renormalization-group recursion relations of the Migdal-Kadanoff approximation are identical to those of an exact solution of a hierarchical lattice \cite{BerkerOstlund,Kaufman1,Kaufman2}.  For recent works using hierarchical lattices, see \cite{Jiang,Derevyagin2,Chio,Teplyaev,Myshlyavtsev,Derevyagin,Shrock,Monthus,Sariyer,Ruiz,Rocha-Neto,Ma,Boettcher5}

\begin{figure}[ht!]
\centering
\includegraphics[scale=0.25]{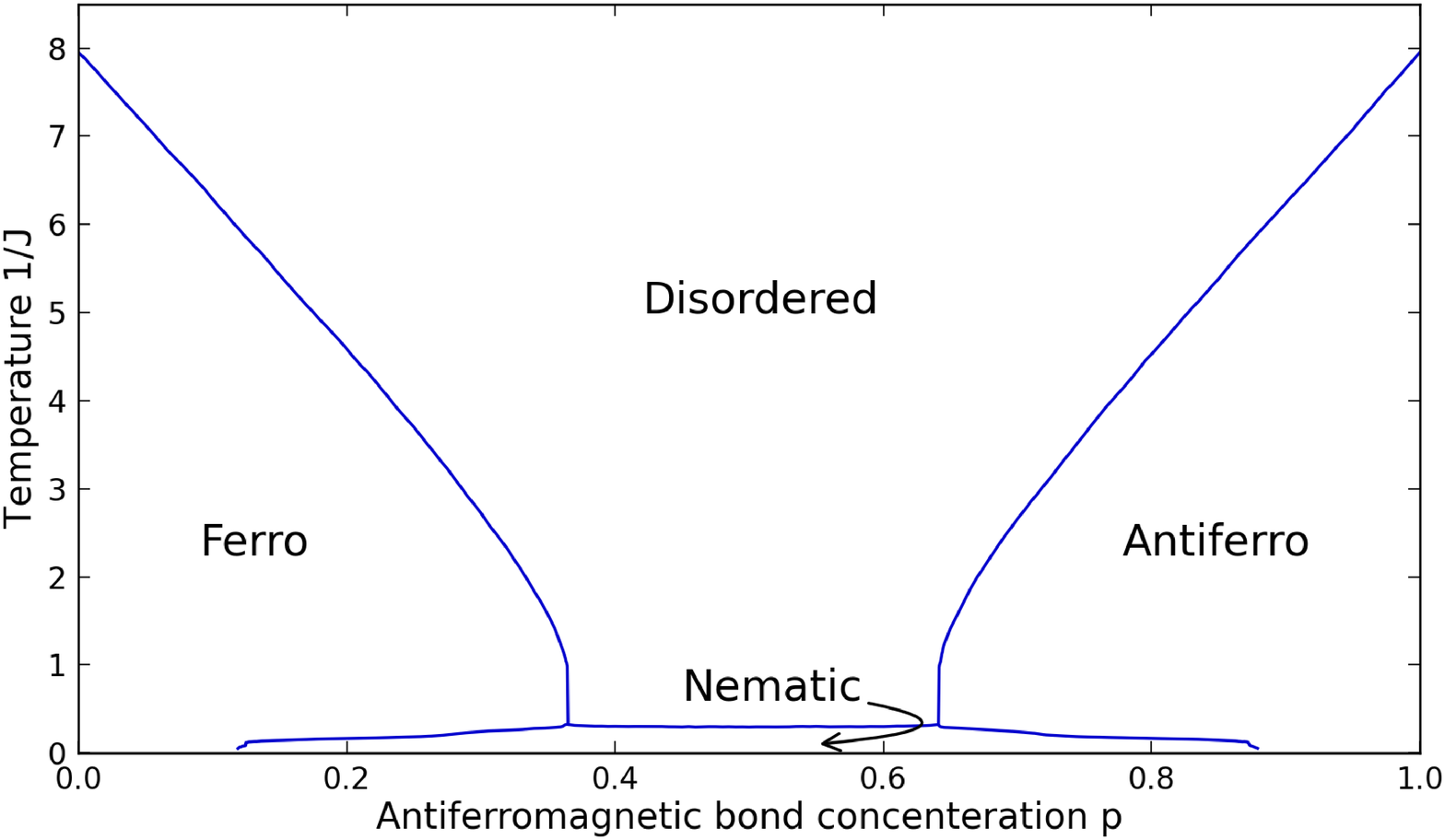}
\caption{Calculated phase diagram of the classical Heisenberg spin-glass system in $d=3$.  The phase diagram shows no spin-glass phase, but at low temperatures an extended nematic phase where the spins globally align along a spontaneously chosen axis irrespective of direction.}
\end{figure}

This simple renormalization-group transformation has been widely successful on different systems: the lower-critical dimension $d_c$ below which no ordering occurs has been correctly determined as $d_c=1$ for the Ising model \cite{Migdal,Kadanoff}, $d_c=2$ for the XY \cite{Jose,BerkerNelson} and Heisenberg \cite{Tunca} models, and the presence of an algebraically ordered phase has been seen for the XY model \cite{Jose,BerkerNelson,Sariyer}. In $q$-state Potts models, the number of states $q_c$ for the changeover from second-order to first-order phase transitions has been correctly obtained in $d=2$ and 3.\cite{Devre} In systems with frozen microscopic disorder (quenched randomness), $d_c=2$ has been determined for the random-field Ising \cite{Machta,Falicov} and XY models \cite{Kutay}, and the non-integer value of $d_c=2.46$ for the Ising spin-glass \cite{Parisi,Amoruso,Bouchaud,Boettcher,Demirtas,Parisi2,Atalay}. Also under the Migdal-Kadanoff approximation, the chaotic nature of the Ising spin-glass phases \cite{McKayChaos,McKayChaos2,BerkerMcKay} has been obtained and Lyapunov exponentwise quantitatively analyzed, both for quenched randomly mixed ferromagnetic-antiferromagnetic spin glasses \cite{Ilker1,Ilker2,Ilker3} and right- and left-chiral (helical) spin glasses \cite{Caglar1,Caglar2,Caglar3}.

\section{Migdal-Kadanoff Renormalization Group for the Heisenberg Model with Non-Uniform Interactions}

The algebra of the Migdal-Kadanoff transformation for discrete spin systems such as Ising, Potts, and clock models is quite simple.  The transformation for the three-component classical Heisenberg model, with each spin having two continuously varying sterangles, has only been recently achieved \cite{Tunca}, for systems without randomness, and is not simple.  Here we generalize this renormalization-group transformation to quenched random systems.

\begin{figure}[ht!]
\centering
\includegraphics[scale=0.24]{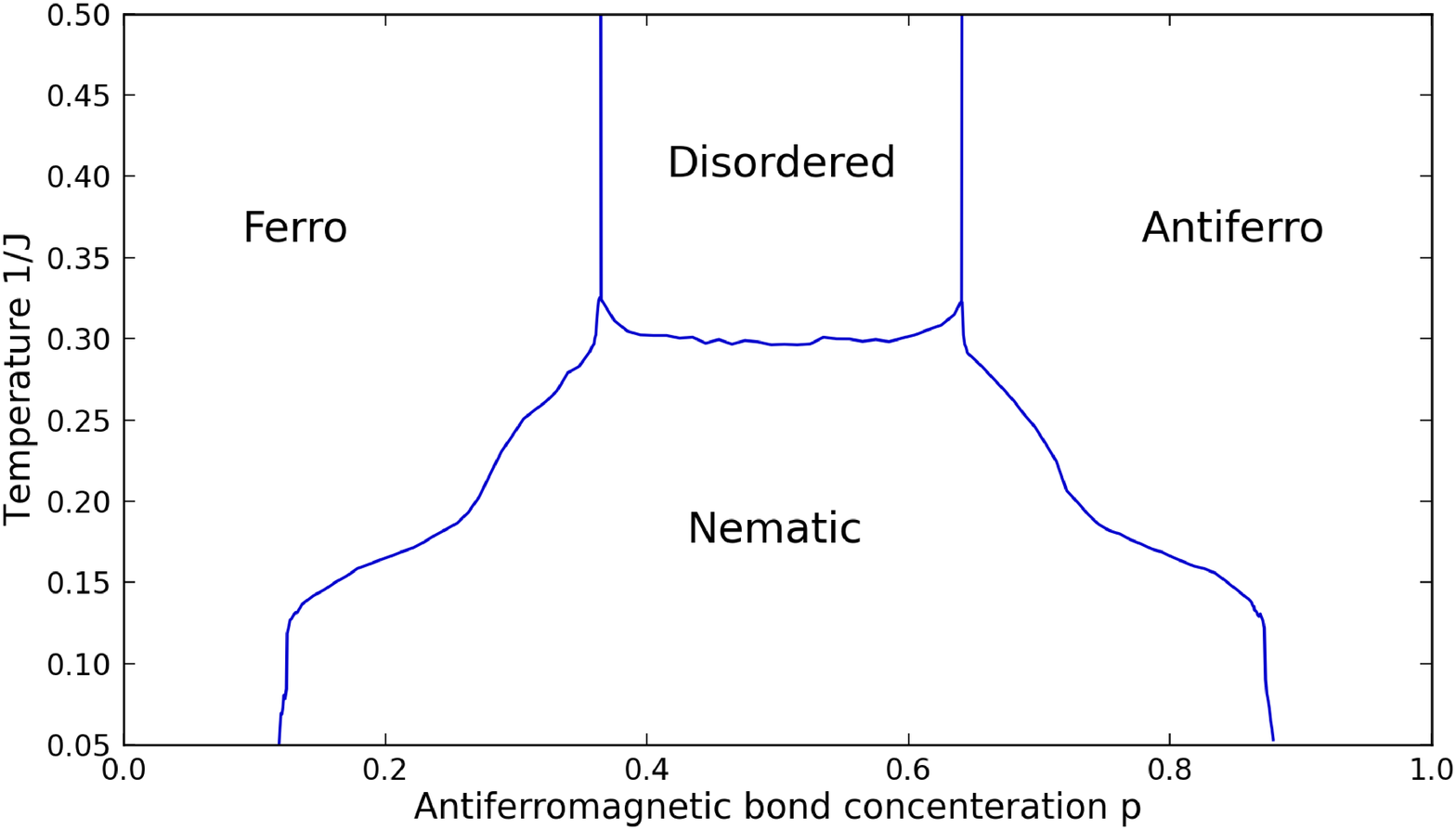}
\caption{Low-temperature portion of the calculated phase diagram of the classical Heisenberg spin-glass system in $d=3$.  As more fully seen here, the phase diagram shows no spin-glass phase, but at low temperatures an extended nematic phase where the spins globally align along a spontaneously chosen axis irrespective of direction.}
\end{figure}

In the first, bond-moving, step (Fig.1) of the Migdal-Kadanoff transformation,
\begin{equation}
\tilde{u}_{i'j'}(\gamma) = u_{i_1j_1}(\gamma)u_{i_2j_2}(\gamma),
\end{equation}
where $u_{ij}(\gamma) = e^{- \beta {\cal H}_{ij}(\vec s_i,\vec s_j)}$ is the exponentiated nearest-neighbor Hamiltonian between sites $(i,j)$ and $\gamma$ is the angle between the spherical unit vectors $(\vec s_i,\vec s_j)$. The tilda denotes bond-moved. Using the Fourier-Legendre series,
\begin{equation}
u_n(\gamma) = \sum_{l=0}^{\infty}{\lambda_l^{(n)} P_l(\cos(\gamma))},
\end{equation}
where $n$ denotes ${i_nj_n}$, with the expansion coefficient $\lambda_l^{(n)}$ evaluated as
\begin{equation}
\lambda_l^{(n)} = \frac{2l+1}{2}\int_{-1}^{1}{u_n(\gamma)P_{l}(\cos(\gamma))\,d(\cos(\gamma))}.
\end{equation}
Thus, for the left side of Eq.(2),
\begin{multline}
\tilde{\lambda}_{l}^{(n')}=\frac{2l+1}{2}\int_{-1}^{1}u_{n_1}(\gamma)u_{n_2}(\gamma)P_{l}(\cos{\gamma})\,d(\cos{\gamma}) =\\
 \frac{2l+1}{2}\sum_{l_1=0}^{\infty}\sum_{l_2=0}^{\infty}\lambda_{l_1}^{(n_1)}\lambda_{l_2}^{(n_2)} \,\cdot\\
\int_{-1}^{1}P_{l_1}(\cos{\gamma})P_{l_2}(\cos{\gamma})P_{l}(\cos{\gamma})\,d(\cos{\gamma})\\
=\sum_{l_1=0}^{\infty}\sum_{l_2=0}^{\infty}\lambda_{l_1}^{(n_1)}\lambda_{l_2}^{(n_2)}{\langle l_1 l_2 0 0 | l_1 l_2 l 0\rangle}^2,
\end{multline}
where the bracket notation is the Clebsch-Gordan coefficient with the restrictions $l_1+l_2+l=2s, s\in \mathbf{N} $; $|l_1-l_2|\leq l\leq |l_1+l_2|$.

In the second, decimation, step of the Migdal-Kadanoff transformation, a decimated bond is obtained by integrating over the shared spin of two bonds,
\begin{multline}
u'_{13}(\gamma_{13}) = \int \tilde{u}_{12}(\gamma_{12}) \tilde{u}_{23}(\gamma_{23}) \frac {d\vec s_2}{4\pi} =    \\
=\sum_{l_1=0}^{\infty}\sum_{l_2=0}^{\infty}\int \tilde{\lambda}_{l_1}^{(n_1)}\tilde{\lambda}_{l_2}^{(n_2)} P_{l_1}(\cos\gamma_{12}) P_{l_2}(\cos\gamma_{23}) \frac {d\vec s_2}{4\pi}.
\end{multline}
The prime denotes renormalized.  Expressing the Legendre polynomials in terms of spherical harmonics,
\begin{multline}
= \sum_{l_1=0}^{\infty}\sum_{l_2=0}^{\infty}\sum_{m_1=-l_1}^{l_1}\sum_{m_2=-l_2}^{l_2}\tilde{\lambda}_{l_1}^{(n_1)}\tilde{\lambda}_{l_2}^{(n_2)}\frac{(4\pi)^2}{(2l_1+1)(2l_2+1)}\,\cdot\\
\int Y_{l_1}^{m_1}(\vec{s_1})Y_{l_1}^{m_1*}(\vec{s_2})Y_{l_2}^{m_2}(\vec{s_2})Y_{l_2}^{m_2*}
(\vec{s_3})\frac {d\vec s_2}{4\pi},
\end{multline}
evaluating the integral and summing over the resulting delta functions,
\begin{multline}
= \sum_{l_1=0}^{\infty}\sum_{m_1=-l_1}^{l_1} \tilde{\lambda}_{l_1}^{(n_1)}\tilde{\lambda}_{l_1}^{(n_2)} \frac{4\pi}{(2l_1+1)^2}Y_{l_1}^{m_1}(\vec{s_1})Y_{l_1}^{m_1*}(\vec{s_3}),
\end{multline}
due to occcuring Dirac delta functions. Rearranging the spherical harmonics back to Legendre polynomials and combining with Eq.(5),
\begin{multline}
\lambda_l^{(n')} = \frac{1}{(2l+1)}\big( \sum_{l_1=0}^{\infty}\sum_{l_2=0}^{\infty}\lambda_{l_1}^{(n_1)}\lambda_{l_2}^{(n_2)}{\langle l_1 l_2 0 0 | l_1 l_2 l 0\rangle}^2\big)\,\cdot\\
\big(\sum_{l_3=0}^{\infty}\sum_{l_4=0}^{\infty}\lambda_{l_3}^{(n_3)}\lambda_{l_4}^{(n_4)}{\langle l_3 l_4 0 0 | l_3 l_4 l 0\rangle}^2\big),
\end{multline}
the full recursion relations of the renormalization-group are obtained. The bond-moved $\tilde{\lambda}$ were substituted from Eq.(5).  Thus, the renormalization-group transformation is in terms of the Fourier-Legendre coefficients $\lambda_l^{(n')}(\{\lambda_l^{(n_i)}\})$.  We have kept up to $l=25$ in our numerical calculations of the trajectories.

\begin{figure}[ht!]
\centering
\includegraphics[scale=0.25]{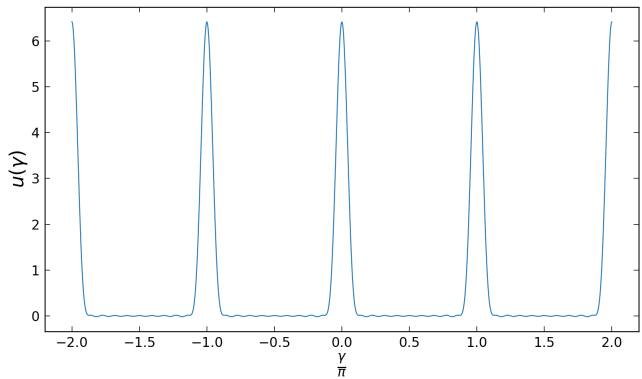}
\caption{The fixed-point exponentiated potential $u(\gamma)$ at the renormalization-group sink of the nematic phase.  The neighboring spins align (nearest-neighbor angle $\gamma=0)$ or anti-align ($\gamma=\pi$), creating the nematic phase with a global spontaneous alignment axis along which both spin directions occur. All points in the nematic phase flow, under repeated renormalization-group transformations, to this sink which epitomizes the ordering of this phase. This potential function, in terms of the nearest-neighbor angle $\gamma$, is reconstructed from the Fourier-Legendre coefficients at the renormalization-group sink.}
\end{figure}

\section{Migdal-Kadanoff Renormalization Group for the Heisenberg Model with Quenched Randomness}

Having derived the renormalization-group transformation for non-uniform nearby interactions, we can now proceed with the solution of the quenched random problem of the spin-glass Heisenberg system in $d$ dimensions, exactly on hierarchical lattices and Migdal-Kadanoff approximately on hypercubic lattices.  We start with 30,000 local ferromagnetic and antiferromagnetic interactions as dictated by the antiferromagnetic probability $p$ given after Eq.(1).  We randomly select from this group to generate 30,000 new interactions. Remembering that for each interaction, 25 Fourier-Legendre coefficients are kept, this is a gigantific calculation. In order to conserve the ferromagnetic-antiferromagnetic symmetry of the system, the length rescaling factor of $b=3$ is chosen.  In the bond-moving step, $b^{d-1}$ interactions are moved onto one interaction.  In the decimation step, $b$ interactions in series are decimated into one.

The renormalization-group trajectories (of sets of 30,000 interactions) are effected by repeated applications of the above transformation.  The initial points of these trajectories are obtained from the Hamiltonian in Eq.(1), which can be written as
\begin{equation}
-\beta \mathcal{H}=\sum_{<ij>}J_{ij}{\vec s_i\cdot \vec s_j}= \sum_{<ij>}J_{ij}{\cos{\gamma_{ij}}}.
\end{equation}
Using the plane-wave expansion for the term in the partition function involving the two spins,
\begin{equation}
e^{J\cos{\gamma}}=\sum_{l=0}^{\infty}{(2l+1)i^l j_l(-iJ)P_l(\cos{\gamma})}=\sum_{l=0}^{\infty}{\lambda_l P_l(\cos{\gamma})},
\end{equation}
where $j_l(-iJ)$ is a spherical Bessel function and $P_l(\cos{\gamma})$ is a Legendre polynomial.

With no approximation, after every decimation and after setting up the initial conditions, the coefficients $\{\lambda_l\}$ are divided by the largest $\lambda_l$.  This is equivalent to subtracting a constant term from the Hamiltonian and prevents numerical overflow problems in flows inside the ordered phases.

\section{Nematic Phase: Global Alignment Spontaneously Generated from Spin-Glass Disorder}

Under repeated applications of the renormalization-group transformation of Eq.(9), the Fourier-Legendre coefficients flow to a stable fixed point, which is the sink of a thermodynamic phase.  The sinks of the disordered phase and the ferromagnetic phase have been discussed and analyzed elsewhere \cite{Tunca}.  The sink of the antiferromagnetic phase is identical to the sink of the ferromagnetic phase, except that the sharp central peak is at nearest-neighbor angle $\gamma = \phi$.

\begin{figure}[ht!]
\centering
\includegraphics[scale=0.5]{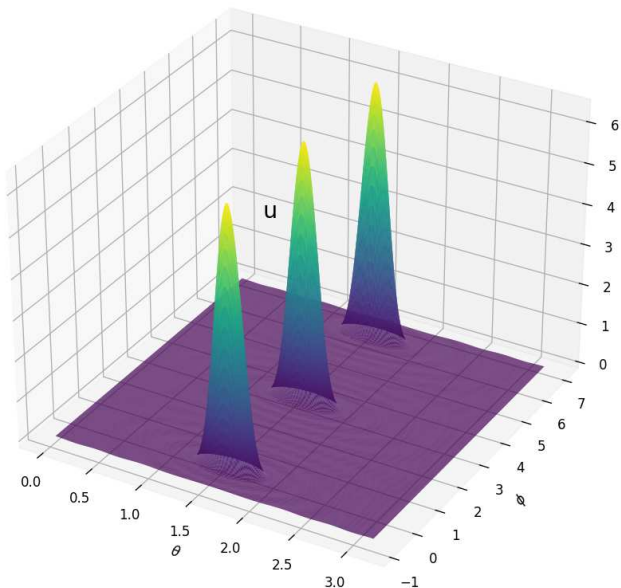}
\caption{The fixed-point exponentiated potential $u(\gamma)$ of the sink of the nematic phase of the $d=3$ classical Heisenberg spin-glass system. This potential function, in terms of the spherical coordinate angles $\theta$ and $\phi$ of one spin with respect to the other, is reconstructed from the Fourier-Legendre coefficients at the sink.}
\end{figure}

For $d=3$ for the classical Heisenberg spin system, a new phase occurs in the low-temperature quenched-disorder region of the phase diagram, as seen in Figs. 2 and 3, where the spin-glass phase is for the Ising system.  The sink of this phase is shown in Figs. 4 and 5.  Fig. 4 shows, at the sink, the exponentiated nearest-neighbor Hamiltonian $u_{ij}(\gamma) = e^{- \beta {\cal H}_{ij}(\vec s_i,\vec s_j)}$ between sites $(i,j)$ versus the angle $\gamma$ between the spherical unit vectors $(\vec s_i,\vec s_j)$. Fig. 5 shows, at the sink, the exponentiated nearest-neighbor Hamiltonian $u_{ij}(\gamma)$ versus the angles $\theta$ and $\phi$ between $(\vec s_i,\vec s_j)$. It is thus seen that the neighboring spins align (nearest-neighbor angle $\gamma=0)$ or anti-align ($\gamma=\pi$), globally creating the nematic phase, where a spontaneous alignment axis along which both spin directions occur. All points in the nematic phase flow, under repeated renormalization-group transformations, to this sink which epitomizes the global ordering of this phase.

It is seen that, in the Heisenberg spin-glass system, at low temperature, this nematic phase extends wider, from $p=0.12$ to 0.88, as compared with the identically placed spin-glass phase of the Ising spin-glass system.  A similar widening, from $p=0.24$ to 0.76 to essentially $p=0$ to 1 is seen \cite{Gulpinar} in the Ising spin-glass phase, when thermal vacancies are included, making domain flipping more favorable, thus eating into the ferromagnetic (and antiferromagnetic) phases without loosing order.  A similar mechanism may be in effect in the present case, with the continuously varying directions of the Heisenberg spins making domain flipping more favorable.

The nematic phase of the classical Heisenberg spin-glass system is also calculated in $d=2.26, 2.46, 2.63, 2.77, 2.89$ dimensions and in dimensions $d \geq 3$.  Our calculated transition temperatures, for 24 dimensions in $2 < d \leq 4$, are shown in Fig. 6.  No nematic (or ferromagnetic \cite{Tunca}) phase occurs in $d=2$, which is expected.

\begin{figure}[ht!]
\centering
\includegraphics[scale=0.23]{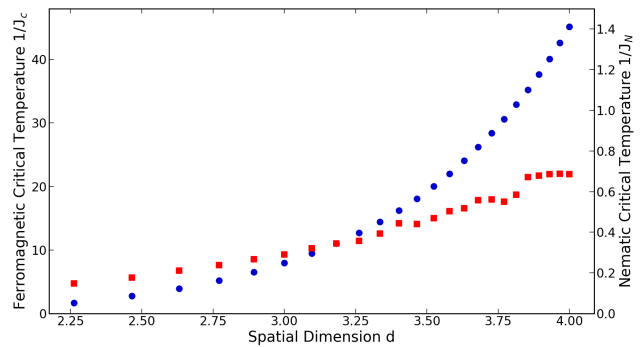}
\caption{Calculated transition temperatures for the nematic phase (at $p=0.5$), squares and right vertical scale, and for the ferromagnetic phase (at $p=1$), circles and left vertical scale, for 24 dimensions in $2 < d \leq 4$.  No nematic (or ferromagnetic \cite{Tunca}) phase occurs in $d=2$, which is expected.}
\end{figure}

\section{Conclusion}
We have solved the classical Heisenberg spin-glass system by renormalization-group theory.  In $d > 2$, in this system with quenched local randomness, a low-temperature phase with global order, in the form of a spontaneously chosen spin easy-axis, irrespective of spin direction.  Thus, a nematic phase occurs in the Heisenberg spin system with competing ferromagnetic and antiferromagnetic interactions.

\begin{acknowledgments}
We grateful to E. Can Artun for useful conversations.  Support by the TEBIP High Performers Program of the Board of Higher Education of Turkey and by the Academy of Sciences of Turkey (T\"UBA) is gratefully acknowledged.
\end{acknowledgments}

\end{document}